\documentclass[conference]{IEEEtran}

\usepackage[numbers]{natbib}
\usepackage[noend]{algpseudocode}
\usepackage{oz}
\usepackage{amsfonts}

\usepackage[ruled,linesnumbered,vlined]{algorithm2e}

\usepackage[below]{placeins} 
\usepackage{caption} 
\usepackage{subcaption} 
\usepackage{hyperref}
\usepackage{ mathrsfs }

\usepackage{color}
\usepackage{graphicx}
\usepackage{multicol}

 \makeatletter
\def\BState{\State\hskip-\ALG@thistlm}
\makeatother
 
 \title{A Formal Specification of a Data  Model for Malaria Surveillance in the Developing World}
 
 \author{\IEEEauthorblockN{Emmanuel Tuyishimire
}

\IEEEauthorblockA{ College of Science and Technology,
University of Rwanda, Kigali, Rwanda (tuyishimire@aims.ac.za)}
}

\begin{document}


%


\maketitle
\pagenumbering{gobble}

\begin{abstract}

 The fourth Industrial Revolution(4IR), together with the COVID-19 pandemic have made a loud call for digitizing diagnosis processes. The world is now convinced that  it is imperative to digitize the diagnosis of long  standing diseases such as malaria for more efficient treatment and control. It has been seen that malaria control would benefit a lot from digitising its diagnosis processes such as data gathering. We propose, in this paper, the architecture of a digital data collection system and how it is used to gather data for malaria awareness. The system is formally specified using Z notation, and  based on the capability of the system, possible malaria determinants are defined and their retrieving mechanisms are discussed.
 
\end{abstract}


\section{Introduction}\label{intro}

During the the Fourth Industrial Revolution (4IR) era, 2019 has marked a year where COVID-19 pandemic surfaced and various processes had to be done remotely. Technology has been widely used to mitigate many issues caused by the pandemic, while contributing to the pandemic control and treatment. This has left an inspiration on how the treatment of existing diseases may be digitized for a more efficient diagnosis. This paper revisits the malaria monitoring and treatment digital system and explore how this disease may be more accurately predicted.

Malaria is a long-standing diseases caused by a plasmodium parasite transmitted through the biting of an infected female Anopheles mosquito. The severity of such a disease depends on the involved types of transmitted parasites, which may be  \textit{Plasmodium} falciparum, \textit{Plasmodium} vivax, \textit{Plasmodium} ovale curtisi, \textit{Plasmodium} ovale wallikeri, \textit{Plasmodium} malariae and the very rare \textit{Plasmodium} knowlesi, as reported in \cite{lalremruata2017species}.

Malaria is one of the most life-threatening diseases which is, however, curable. In 2021, 241 millions cases have been recorded and a total of 627000 malaria-deaths have been estimated. Africa has been the most affected region.  $90\%$ of the total global cases and $90\%$ of the malaria death have been found in Africa, and $77\%$ of the death were children under 5 years old \cite{jagannathan2022malaria}. On the other hand,  since 2009, many African countries  have required that all malaria cases need to be confirmed through
parasitology tests,   using the light microscopy. However, microscopy has later been identified as  a time-consuming process whose efficacy varies based on many factors including the viral load, the course of infection, and the skill of the technician. While skilled technicians can detect malaria with a Plasmodium density of 30 parasites per micro litre of smeared blood, novice ones may require up to 100 parasites to be present \cite{dahal2021challenges}. With such a wide variance, it is unsurprising that a meta-analysis on the comparative effectiveness of different malaria parasitology tests found that light microscopy missed up to $50\%$ of cases (diagnosed via real-time Polymerase Chain Reaction Tests i.e. PCR) \cite{okell2009submicroscopic}.

Despite this, using microscopy on stained blood smears remains the best choice for malaria diagnosis in resource-limited environments for several reasons:
\begin{itemize}
    \item It is less costly than PCR and rapid diagnostic tests (RDTs).
    \item It can be more accurate than RDTs.
    \item It can provide an actual count of the number of Plasmodium species in a blood smear, potentially indicating the degree of severity of the illness and allowing more accurate treatment from the onset.
    \item It can identify the different species of Plasmodium, unlike RDTs which only confirm the presence or absence of Plasmodium. This can not only allow the appropriate treatment to be provided, but also provide real-time data on the trends of malaria cases caused by different Plasmodium species to be collected. 
\end{itemize}

Furthermore, the tool needs to be complemented by advanced storage and computation capabilities to respond to the prediction and reinforcement issues. It is important to have a well structured central system where such test datasets are stored and well managed. We are in the era where advanced communication and computer assistance has risen the necessity to describe the information systems using formal languages such as the Z notation \citep{spivey1992z}.

 On the other hand, collecting malaria data has been a subject of research. It is nowadays possible to capture the macroscopic images, analyse and use it to digitally test malaria \cite{okell2009submicroscopic}. This approach can benefit from existing data collection and  transport models such as \cite{tuyishimirecooperative,jibreel2022enhanced, tuyishimire2019clustered} to be aggregated on a national or regional level, where they can further be processed for a prediction, reinforcement and visualisation. However the formally structured aggregation of such delivered data is still a subject of research.

Formal methods have been highly recommended  to specify, develop and verify complex information systems \citep{wing1990specifier}. The healthcare document sharing has been specified using Z notation in \citep{ali2018z}, and the access control mechanisms have been formally discussed. On the other hand, formal methods help in accurately specifying systems whose error and discrepancies are identified during early phases of the software development process. This has motivated the authors in \citep{azeem2014specification} to propose an e-health system, using Z notation. These works are too generic for tracking the treatment of a disease like malaria.

Various medical tools and operations have been developed using formal methods \citep{bonfanti2018systematic}. These include Hemodialysis Machine, Pacemaker, Infusion pump, Medical image processing, Stereoacuity test, ECG (Electrocardiography), and many others. However the survey has not identified any formal methods-based description of a central data gathering system to complement the diseases treatments.

On the other hand, various centralised information systems has been proposed for monitoring purpose. In \citep{bagula2021cyber},  a Cyber-Physical system which is dependable for Internet of Things (IoT) applications, has been proposed. For health applications, a cyber-healthcare kiosk model has been proposed in \cite{bagula2018cyber}, to apply IoT in automatically determining health condition in the developing world.  The World Health Organisation (WHO) has illustrated a digital system model for malaria surveillance in  \citep{who2018malaria}. These systems have been proven efficient but  their formal specification is, to our best knowledge,still pending.

In this paper, we use Z notation to formally specify a information system for malaria data gathering, aiming its surveillance. The architecture of the data collector is described and formally specified. The models for retrieving various malaria determinants are discussed. We are aware of the fact that several systems such as e-commerce, schedules, remuneration, task assignment, ect., are to, cooperatively, work together with the data gathering system. In this work we obstruct these additional systems and focus on the malaria treatment system. We are taking advantage of how Z notation is very flexible when it comes with integration and scalability.

The rest of this paper is organise as follows.
Section \ref{pro} covers the architecture of the proposed digital system, in Section \ref{dis} the proposed epidemic model is detailed, the system performance is studied in Section \ref{ret} and Section \ref{con} concludes the paper, while suggesting future works.

\section{Proposed architecture}\label{pro}

In this section, we propose the architecture of a data management system aiming the efficient data collection and process for improving the aspects of malaria forecasting. Here, we are inspired by a Cyber-Physical system which is dependable for Internet of Things (IoT) applications, as proposed in \cite{bagula2021cyber}. The system consists of an IoT system which enables data collection for the system forecasting purpose.

On the other hand, the World Health Organisation (WHO) has illustrated a digital system model for malaria surveillance in  \cite{who2018malaria}. Here, a data and information flow systems has been briefly described for national decision making purpose. However, the mechanisms and technology to enable the system, in the African context would complement the proposed work.

\subsection{System Architecture}

\begin{figure}[ht!]
\centering
\includegraphics[scale=0.3]{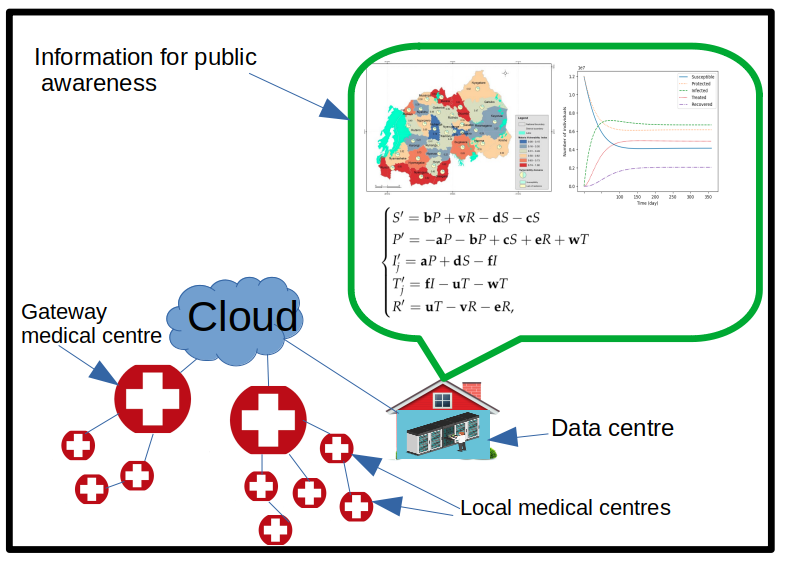}
\caption{Proposed model}
\label{model}
\end{figure}

\begin{itemize}
\item [$\blacktriangleright$] Local medical centres:these are ordinary medical centres where people go for malarial treatment. They get tested and receive medication there. Medical centres  can also communicate with each other using various ways such as aerial \cite{tuyishimire2022trajectory, tuyishimire2016internet, bagula2016internet, tuyishimirecooperative, ismail2018internet,ismail2018generating}, terrestrial roads and electronic communication channels \cite{mauwa2019optimal, tuyishimire2020formal}.

\item [$\blacktriangleright$] Gateway medical centres: these have same properties  as local medical centres but they additionally have the capability to push the data in the central system, where they may be processed further.
\item [$\blacktriangleright$] Cloud: it is the central system where the data are gathered , preprocessed and shared to relevant stakeholders.
\item [$\blacktriangleright$] Data centre: The data centre is the physical institution which deals with all aspects of data management.
\item [$\blacktriangleright$] Information for public awareness: these are information obtained by processing the gathered data on the cloud.

\end{itemize}
  
  \subsection{System specification} 
%
To formally specify the system, we first identify the basic observables for malaria data processes. We then define complex observables using Z schemas which are translated and explained. We are inspired by a dependable data management model proposed in \citep{malaria2022species}.

\subsubsection{The identifier $NAME$} The type $NAME$ is used to identify individuals, places and items. The set $NAME$ of identifiers may be extended to cover the case of non identifiability as follows.

$$NAME_-= NAME \cup \{ -\}$$

where the symbol $-$  stands for a non-existing or undefined element. Representing a non-existing helps in quantitative analysis such as counting.

\subsubsection{Malaria compartments $S,\ P,\ I,\ T,\ R$}

We assume 5 different malaria statuses, namely, Susceptible, Infected, Recovered, Protected,  Treated as explained as follows.

\begin{enumerate}

\item Susceptible people ($S$): The set $\mathcal{S} $ consists of people who are exposed to the disease. The cardinality of the set is defined by a column matrix $S$ whose $j^{th}$ entries corresponds to the number of susceptible humans of the $j^{th}$ type of the disease.

\item Protected ($P$): Some of susceptible people may consider to take some protective measures such as IRS and or ITN to be immune for a period of time and once such a period elapses, some of these people may be susceptible again. The cardinality for such a set is expressed by the following column matrix.

\item Infected people ($I $): it is the set of people who have been contaminated. These are the people who have been either susceptible or protected and their protection period elapsed.

\item Treated people ($T $): this  is the set of people who have been contaminated and are consuming medicines. It is here assumed that such people cannot be exposed to the same disease for with they are being taking medicines. After treatment, people may immediately take protection measures for them to remain protected for some time. The number of Treated people may be expressed in terms of the following column vector.

 It is important to  note that the malaria treatment is not always variant oriented. That is, various malaria variants my be treated at the same time even if one variant has been detected.

\item Recovered people ($R $): it is the set of people who have been successfully cured because of the medical treatment or traditional healing. Here, it is important to mention that in deep villages, some people may use some plants and successfully cure Malaria. In this case we assume that all Malaria variants have been cured. Such treatment may be recorded using local health advisors, who are nationally recognised ("Abajyanama b'ubuzima" for example, in the case of Rwanda). These people may become protected for a small period, and later, they may consider taking protection measures. Furthermore, such people may become exposed to the disease again. The cardinality of Recovered people is expressed by the following column matrix.

\end{enumerate}

\subsubsection{Malaria status $STATUS$}

It is important to mention that malaria is caused by various kinds of plasmodium parasites. 

Assume $X$ to be the set of all available parasites. The corresponding ordered set is denoted by $seqX$. The ordered set $seqX$ is mapped to the tuple whose size is the same as the cardinality of the set, which contain binary values. We will refer to this tuples as the Disease Binary Code (DBC). If the $j^{th}$ entry of the DBC is 1, then the corresponding parasite in $seqX$ is observable and if  $j^{th}$ entry is  0, the corresponding parasite is not.

The malaria status of an individual is determined as shown in Schema $STATUS$.

The status is determined by the observed conditions of a person with regards to all the available parasites. This is determined by the DBC for each observable compartment. In the case the $j^{th}$ entry (denoted by $s_j$), in the DBC for Susceptible $STATUS.s$, is equal to one, then for  the other compartments protected $STATUS.p$, infected $STATUS.i$, treated $STATUS.t$ and recovered $STATUS.r$, the $j^{th}$ entry is zero. This is also the same for all other compartments. This means that a person cannot be in more than one compartments at the same time.

Note that the first five predicates and the last five ones show that the updated status satisfies the predicates for the type $STATUS$. Furthermore, the schema shows that updating the stratus means overwriting the existing data available in the previous one. 
\begin{class}{STATUS}
\small
\begin{state}
s, p, i, t, r: ran(seqX \rightarrow \{0,1\}^{\#X})\\
\where
s_j=1 \Rightarrow p_j= i_j= t_j= r_j=0\\
p_j=1 \Rightarrow s_j= i_j= t_j= r_j=0\\
i_j=1 \Rightarrow s_j= p_j= t_j= r_j=0\\
t_j=1 \Rightarrow s_j= p_j= i_j= r_j=0\\
r_j=1 \Rightarrow s_j= p_j= i_j= t_j=0\\
\end{state}\\
\begin{op}{updateSTATUS}
\Delta (s, p, i, t, r)\\
s?, p?, i?, t?, r?: ran(seqX \rightarrow \{0,1\}^{\#X})\\
\where
s'=s?\\
p'= p?\\
i'= i?\\
t'= t?\\
r'= r?\\
s?_j=1 \Rightarrow p?_j= i?_j= t?_j= r?_j=0\\
p?_j=1 \Rightarrow s?_j= i?_j= t?_j= r?_j=0\\
i?_j=1 \Rightarrow s?_j= p?_j= t?_j= r?_j=0\\
t?_j=1 \Rightarrow s_j= p?_j= i?_j= r?_j=0\\
r?_j=1 \Rightarrow s?_j= p?_j= i?_j= t?_j=0\\
\end{op}
\end{class}

\subsubsection{Physical address $ADDRESS$} This is the full description of the physical address of an individual or institution. Recording this would help in regional based studies and investigations. In various developing countries, the address is determined by the province $ADDRESS.province$, sector $ADDRESS.sector$ and cell $ADDRESS.cell$. Notice that the street address, house number and postal address are not necessarily right options to define the addresses in the developing world.

It is also important to mention that names for various ADDRESS observables may be the same, and this is the case of Rwanda.

\begin{schema}{ADDRESS}
province, district, sector, cell: NAME
\end{schema}

\subsubsection{Electronic address $SNUMBER$}
The set of all medical centres' electronic identifiers (i.e. addresses) is denoted by $IP$. We extend the set $IP$ to $$IP_-=IP \cup \{-\}$$ where,  $-$ stands for a non existing or undefined address.

\subsubsection{The user $USER$}
\begin{schema}{USER}
name: NAME\\
address: ADDRESS\\
status: STATUS
\end{schema}
The type $USER$ represents a person who needs malaria based services. The user is determined by his/her name $USER.name$, address $USER.address$ and malaria status $USER.status$.

\subsubsection{Staff number $SNUMBER$}
The staff number may be a number reflecting more information such as the starting year, the details of the medical centre, age , gender, etc., depending on which details would inform the malaria treatment. 
\subsubsection{The doctor $DOCTOR$}

\begin{schema}{DOCTOR}
name: NAME\\
stn:SNUMBER \\
address: ADDRESS
\end{schema}

The type $DOCTOR$ represents the person who is ready to provide malaria based services. This type is determined by the name $DOCTOR.nam$, staff number $DOCTOR.stn$ and the address $DOCTOR.address$.

 \subsubsection{The medical centre $MEDCENTRE$}
 
\begin{class}{MEDCENTRE}
\small
\begin{state}
users:  \mathbb{P} USER\\
doctors:\mathbb{P} DOCTOR\\
location: ADDRESS\\
ip: IP_-
\end{state}\\
\begin{op}{addUSER}
\Delta users\\
user?:USER
\where
users'=users  \cup \{user?\}
\end{op}\\
\begin{op}{rmUSER}
\Delta users\\
user?:USER
\where
users'=users \setminus \{user?\}
\end{op}\\
\begin{op}{addDOCTOR}
\Delta doctors\\
doctor?:DOCTOR
\where
doctors'=doctors \cup\{doctor?\}
\end{op}\\
\begin{op}{rmDOCTOR}
\Delta doctors\\
doctor?:DOCTOR
\where
doctors'=doctors \setminus \{doctor?\}
\end{op}
\end{class}

The type $MEDCENTRE$ represents the institutions/class where the malaria services are provided.
It consists of a set of users $MEDCENTRE.users$, the set of doctors $MEDCENTRE.doctors$ and  the unique location $MEDCENTRE.location$.

The medical center  may be updated either  by adding or removing the user (see $addUSER$  and $rmUSER$ schema, respectively). A user/patient is removed when s/he has changed the status.

The update may also be done regarding  doctors  as indicated by the 
$addDOCTOR$ and $rmDOCTOR$  schemas.

 \subsubsection{The data centre $DATACENTRE$}

\begin{schema}{DATACENTRE}
ip: IP_-\\
location: ADDRESS\\
\end{schema}

The type $DATACENTRE$ is the data office where data are delivered aggregated, and preprocessed. In this paper we assume that all data re digitised and hence stored in an infrastructure determined by IP address $DATACENTRE.ip$, and a geographical address $DATACENTRE.location$.

 \subsubsection{The networked medical centres $NMEDCENTRES$}

\begin{class}{NMEDCENTRES}
\small
\begin{state}
medcentres: \mathbb{P}MEDCENTRE\\
roads :MEDCENTRE \leftrightarrow MEDCENTRE\\\
datac: DATACENTRE\\
gateways:: MEDCENTRE
\where
gateways \in medcentres\\
roads \in medcentres \leftrightarrow medcentres\\
ran(roads^{*})=medcentres\\
roads^{\sim}=roads\\
id_{medcentres} \cap roads= \emptyset\\
nbr(n)=roads \limg n\rimg\\
datac \notin nodes\\
\forall m_1,m_2 : medcentres \dot m_1.ip \neq m_2.ip \neq datac.ip\\
\end{state}\\
\begin{op}{addMEDCENTRE}
\Delta medcentres\\
medcentre?:MEDCENTRE
\where
medcentres'= medcentres \cup \{ medcentre?\}
\end{op}\\
\begin{op}{rmMEDCENTRE}
\Delta medcentres\\
medcentre?:MEDCENTRE
\where
medcentres'= medcentres \setminus \{ medcentre?\}
\end{op}
\end{class}
The networked medical center consists of the network of all medical centres some of which are gateways. Here, a gateways is defined as a medical center with is directly connected to the data center. This networked system can be update by either removing or adding the medical centres.

\section{Disease modelling and prediction}\label{dis}

\subsection{Spread modelling}

\begin{figure}[ht!]
\centering
\includegraphics[scale=0.35]{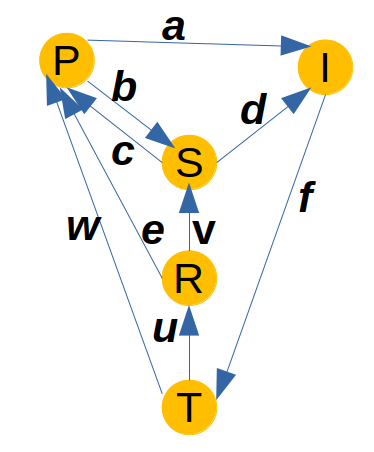}
\caption{Spread model.}
\label{spread}
\end{figure}

Figure \ref{spread} shows the spread rates, expressed as compartment migration rates. For example, the rate of migrating from Susceptible to Infected is denoted by $d$ and from Treated to Protected, the rate is denoted by $w$.

Taking account to all cases discussed in Section \ref{dis}, 
we obtain the following difference equation.
\setlength{\arraycolsep}{0.0em}
\begin{eqnarray}\label{mod} \centering
  S_j'= & - d_jS_j - c_jS_j + b_j P_j + v_jR_j   \nonumber\\
  P_j'= & c_j S_j  -a_j P_j -b_jP_j  + w_jT_j + e_jR_j \nonumber\\
  I_j' = & d_jS_j + a_jP_j -f_jI_j \nonumber\\
  T_j'= & -u_jT_j -w_jT_j+ f_jI_j   \nonumber\\
  R_j' = & u_jT_j - v_iR_i - e_iR_i \nonumber\\
\end{eqnarray}

Note that $S_j$, $P_j$, $I_j$, $T_j$  and  $R_i$  are functions of time $t$ for each variant $j$.
The negative rates in the model represent a decrease, whereas the positive ones represent an increase.

Considering the Hadamard product (element wise multiplication of matrices) Equation \ref{mod} becomes 

\setlength{\arraycolsep}{0.0em}
\begin{eqnarray}\label{mod2} \centering
  S'= &\textbf{b} P + \textbf{v}R - \textbf{d}S - \textbf{c}S  \nonumber\\
  P'= &-\textbf{a} P -\textbf{b}P + \textbf{c} S +\textbf{e}R + \textbf{w}T  \nonumber\\
  I_j' = & \textbf{a}P+ \textbf{d}S -\textbf{f}I \nonumber\\
  T_j'=  &\textbf{f}I -\textbf{u}T -\textbf{w}T  \nonumber\\
  R' =  &\textbf{u}T - \textbf{v}R - \textbf{e}R \nonumber\\
\end{eqnarray}

where the parameters \textbf{a}, \textbf{b}, \textbf{c}, \textbf{d}, \textbf{e}, \textbf{f}, \textbf{u}, \textbf{v} and \textbf{w} are vectors of parameters whose entries correspond to the variants of the disease.


%
%
%
%
%
%
%
%
%
%
%
%

\section{Data retrieval}\label{ret}

\subsection{Regional data retrieval}
Regional distributions such as nation distribution for 
mosquitoes and risk of malaria infection has nationally been proposed in \citep{hakizimana2018spatio} for the case of Rwanda.
\FloatBarrier
\begin{figure}[h]

\small
\centering
\begin{schema}{provinceUSERS[NMEDICENTRES]}
province?: NAME\\
users!: \mathbb{P}USER\\
\where
province? \in \cup\{NMEDICENTRES.medcenters.location.province\}\\
users!=\{NMEDICENTRES.medcenters.users | \\\ \ \ \ \ \ \ \ \ \ \   \forall   NMEDICENTRES.medcenters.location.province \\ \ \ \ \ \ \ \ \  \ \ \ \ = province? \}
\end{schema}
\end{figure}
The same retrieving mechanism may also apply to users from a district,a sector or a cell.

\subsection{Determinants retrieval}
Let $X$ be a set of current parasites and $t$ be the age of the information system (the time ).
Based on a region, say a province, transmission rates may be computed after a period of time. This depends on the current statuses and the previous one. For example the product $ta_i$ is the total number of people who migrated from infected to protected.
\begin{figure}
\small
\centering
\begin{schema}{provinceUSERS[provinceUSERS,X]}
a, b, c, d, e, f, u, v, w: \mathbb{R}^{\# X}\\
province?:NAME
t: \mathbb{R}
\where
province? \in \cup\{NMEDICENTRES.medcentres.location.province\}\\
ta_j= \# \{usr \in provinceUSERS.users!\\
\ \ \ \ \ \ \dot usr.status.i_ j=1 \wedge updateSTATUS^{-1}(usr.status).p_j=1\}\\
tb_j= \# \{usr \in provinceUSERS.users!\\
\ \ \ \ \ \ \dot usr.status.s_ j=1 \wedge updateSTATUS^{-1}(usr.status).p_j=1\}\\
tc_j= \# \{usr \in provinceUSERS.users!\\
\ \ \ \ \ \ \dot usr.status.p_ j=1 \wedge updateSTATUS^{-1}(usr.status).s_j=1\}\\
td_j= \# \{usr \in provinceUSERS.users!\\
\ \ \ \ \ \ \dot usr.status.i_ j=1 \wedge updateSTATUS^{-1}(usr.status).s_j=1\}\\
te_j= \# \{usr \in provinceUSERS.users!\\
\ \ \ \ \ \ \dot usr.status.p_ j=1 \wedge updateSTATUS^{-1}(usr.status).r_j=1\}\\
tf_j= \# \{usr \in provinceUSERS.users!\\
\ \ \ \ \ \ \dot usr.status.t_ j=1 \wedge updateSTATUS^{-1}(usr.status).i_j=1\}\\
tu_j= \# \{usr \in provinceUSERS.users!\\
\ \ \ \ \ \ \dot usr.status.r_ j=1 \wedge updateSTATUS^{-1}(usr.status).t_j=1\}\\
tv_j= \# \{usr \in provinceUSERS.users!\\
\ \ \ \ \ \ \dot usr.status.s_ j=1 \wedge updateSTATUS^{-1}(usr.status).r_j=1\}\\
tw_j= \# \{usr \in provinceUSERS.users!\\
\ \ \ \ \ \ \dot usr.status.p_ j=1 \wedge updateSTATUS^{-1}(usr.status).t_j=1\}\\
\end{schema}
\end{figure}
\FloatBarrier

\section{Conclusion and future work}\label{con}

In this paper a data gathering framework has been formally proposed. Z notation has been used to specify all required data structure. Assuming an epidemic model, the mechanisms for retrieving local disease details has been provided. The system update has been discussed, it has been shown how the compartmental transmission rates may be computed. 

This work is part of a bigger project of digitizing malaria diagnosis process. This consist of gathering data as proposed in this paper, incorporating the convolutional neural network to detect malaria status and related insight will be displayed on a public interface.

\small
\renewcommand{\bibname}{\textit{\large{References}}}
\bibliographystyle{unsrt}
\bibliography{references}
\addcontentsline{toc}{chapter}{References}
\end{document}